\documentclass[twocolumn,resetfootnote]{aastex701}
\usepackage{xspace}
\usepackage{hyperref}

\newcommand{\fluxcgs}{erg~s$^{-1}$~cm$^{-2}$\xspace}

\shorttitle{X-Ray Polarization of Cyg X-2 in the Horizontal Branch}
\shortauthors{Gnarini et al.}

\begin{document}

\title{Discovery of High X-Ray Polarization from the Neutron Star Low-Mass X-Ray Binary Cyg X-2 in the Horizontal Branch}

\author[0000-0002-0642-1135]{Andrea Gnarini}
\affiliation{Dipartimento di Matematica e Fisica, Universit\`{a} degli Studi Roma Tre, Via della Vasca Navale 84, I-00146 Roma, Italy}
\affiliation{NASA Marshall Space Flight Center, Huntsville, AL 35812, USA}
\email{andrea.gnarini@uniroma3.it}

\author[orcid=0000-0002-2381-4184]{Swati Ravi}
\affiliation{MIT Kavli Institute for Astrophysics and Space Research, Massachusetts Institute of Technology, Cambridge, MA 02139, USA}
\email{swatir@mit.edu}

\author[0000-0002-3638-0637]{Philip Kaaret}
\affiliation{NASA Marshall Space Flight Center, Huntsville, AL 35812, USA}
\email{philip.kaaret@nasa.gov}

\author[0009-0009-3183-9742]{Anna Bobrikova}
\affiliation{Department of Physics and Astronomy, FI-20014 University of Turku, Finland}
\email{anna.a.bobrikova@utu.fi}

\author[0000-0002-0983-0049]{Juri Poutanen}
\affiliation{Department of Physics and Astronomy, FI-20014 University of Turku, Finland}
\email{juri.poutanen@utu.fi}

\author[0000-0001-9167-2790]{Sofia V. Forsblom}
\affiliation{Department of Physics and Astronomy, FI-20014 University of Turku, Finland}
\email{sofia.v.forsblom@utu.fi}

\author[0000-0001-9442-7897]{Francesco Ursini}
\affiliation{Dipartimento di Matematica e Fisica, Universit\`{a} degli Studi Roma Tre, Via della Vasca Navale 84, I-00146 Roma, Italy}
\email{francesco.ursini2@uniroma3.it}

\author[0000-0003-1285-4057]{Maria Cristina Baglio}
\affiliation{INAF--Osservatorio Astronomico di Brera, Via Bianchi 46, I-23807 Merate (LC), Italy}
\email{maria.baglio@inaf.it}

\author[0000-0002-4622-4240]{Stefano Bianchi}
\affiliation{Dipartimento di Matematica e Fisica, Universit\`{a} degli Studi Roma Tre, Via della Vasca Navale 84, I-00146 Roma, Italy}
\email{stefano.bianchi@uniroma3.it}

\author[0000-0002-6384-3027]{Fiamma Capitanio}
\affiliation{INAF--Istituto di Astrofisica e Planetologia Spaziali, Via Fosso del Cavaliere 100, I-00133 Roma, Italy}
\email{fiamma.capitanio@inaf.it}

\author[0000-0002-5817-3129]{Massimo Cocchi}
\affiliation{INAF--Osservatorio Astronomico di Cagliari, via della Scienza 5, I-09047 Selargius (CA), Italy}
\email{massimo.cocchi@inaf.it}

\author[0009-0002-1852-7671]{Mar\'ia~Alejandra~D\'iaz~Teodori}
\affiliation{Department of Physics and Astronomy, FI-20014 University of Turku,  Finland}
\affiliation{Nordic Optical Telescope, Rambla José Ana Fernández, Pérez 7, E-38711 Breña Baja, Spain}
\email{alejandra.m.diaz@utu.fi}

\author[0000-0003-1533-0283]{Sergio Fabiani}
\affiliation{INAF--Istituto di Astrofisica e Planetologia Spaziali, Via Fosso del Cavaliere 100, I-00133 Roma, Italy}
\email{sergio.fabiani@inaf.it}

\author[0000-0003-2212-367X]{Ruben Farinelli}
\affiliation{INAF--Osservatorio di Astrofisica e Scienza dello Spazio, Via P. Gobetti 101, I-40129 Bologna, Italy}
\email{ruben.farinelli@inaf.it}

\author[0000-0002-2152-0916]{Giorgio Matt}
\affiliation{Dipartimento di Matematica e Fisica, Universit\`{a} degli Studi Roma Tre, Via della Vasca Navale 84, I-00146 Roma, Italy}
\email{giorgio.matt@uniroma3.it}

\author[0000-0002-0940-6563]{Mason Ng}
\affiliation{Department of Physics, McGill University, 3600 rue University, Montr\'{e}al, QC H3A 2T8, Canada}
\affiliation{Trottier Space Institute, McGill University, 3550 rue University, Montr\'{e}al, QC H3A 2A7, Canada}
\email{masonng@mit.edu}

\author[0000-0003-2609-8838]{Alexander Salganik}
\affiliation{Department of Physics and Astronomy, FI-20014 University of Turku,  Finland}
\email{alsalganik@gmail.com}

\author[0000-0002-7781-4104]{Paolo Soffitta}
\affiliation{INAF--Istituto di Astrofisica e Planetologia Spaziali, Via Fosso del Cavaliere 100, I-00133 Roma, Italy}
\email{paolo.soffitta@inaf.it}

\author[0009-0007-0537-9805]{Antonella Tarana}
\affiliation{INAF--Istituto di Astrofisica e Planetologia Spaziali, Via Fosso del Cavaliere 100, I-00133 Roma, Italy}
\email{antonella.tarana@inaf.it}

\author[0000-0001-5326-880X]{Silvia Zane}
\affiliation{Mullard Space Science Laboratory, University College London, Holmbury St Mary, Dorking, Surrey RH5 6NT, UK}
\email{s.zane@ucl.ac.uk}

% \correspondingauthor{Andrea Gnarini}
% \email{andrea.gnarini@uniroma3.it}

\begin{abstract}
We present results from simultaneous X-ray polarimetric and spectroscopic observations of the bright neutron star low-mass X-ray binary \mbox{Cyg X-2}, performed by the Imaging X-ray Polarimetry Explorer (IXPE) and the Nuclear Spectroscopic Telescope Array (NuSTAR). IXPE detected significant polarization (15$\sigma$) from the source in the 2--8 keV energy band with an average polarization degree (PD) of $4.5\% \pm 0.3\%$ and a polarization angle (PA) of $128\degr \pm 2\degr$ as the source moved along the horizontal branch of its Z-track. The PD increases with energy reaching $9.9\%\pm 2.8\%$ in the 7--8 keV band, with no evidence for energy-dependent variation in the PA. The PA is roughly consistent with previous measurements obtained during the normal and flaring branches and also with the known radio jet axis. From spectropolarimetric analysis, the main contribution to the polarized radiation is due to Comptonized photons, but the polarization is higher than predicted in typical spreading layer geometries. The observed high polarization may be due to a combination of a highly polarized reflected component and a moderately polarized spreading layer on the neutron star surface or produced by electron scattering in an equatorial wind.
\end{abstract}

\section{Introduction}

Low-mass X-ray binaries (LMXBs) hosting weakly magnetized neutron stars (WMNSs) form some of the most extreme environments in astrophysics \citep{Lewin2006}. These objects are among the brightest persistent X-ray sources in the sky, making them excellent subjects for spectroscopic, timing, and polarimetric observational studies. In these systems, the matter from a companion star with a mass not exceeding that of the Sun falls onto a neutron star via Roche-lobe overflow, resulting in the formation of an accretion disk \citep{Frank2002}. The disk is disturbed near the neutron star, forming a boundary layer \citep[BL; see, e.g.,][]{Shakura1988} or a spreading layer \citep[SL; see, e.g.,][]{Inogamov1999} near the NS surface. The emission coming from the BL/SL region can be reflected off the disk (see the review by \citealt{Ludlam2024}) or scattered in a wind above the disk \citep[e.g.,][]{DiazTrigo2016,Ponti2016}. Together with the direct emission of the disk and the BL/SL region, these processes shape the spectrum and polarization of the X-rays emitted from the system. 

The X-ray emission from WMNS binary systems exhibits strong variability over a wide range of timescales. The short time scale variability includes thermonuclear type-I X-ray bursts \citep[e.g.,][]{Lewin1993,Galloway2008} and coherent pulsations \citep[e.g.,][]{Wijnands1998,Patruno2021}, both of which identify the compact objects as neutron stars. On longer time scales, WMNS binaries transition between harder and softer states, with both their luminosity and their spectra changing significantly. Their spectral variations, as depicted in color-color diagrams (CCDs) and hardness-intensity diagrams (HIDs), are used to classify the WMNS-LMXBs into two major classes, atoll and Z-sources \citep[see, e.g.,][]{Hasinger1989}. In particular, Z-sources are characterized by higher luminosity ($> 10^{38}$~erg~s$^{-1}$) and a wide Z-like track in the CCD, consisting of the horizontal branch (HB), normal branch (NB), and flaring branch (FB). The turning points of the Z-track are the hard apex (HA) between the HB and NB and the soft apex (SA) between the NB and FB \citep{Hasinger1989,Blom1993}. Both classes (Z- and Atoll sources) have been extensively studied over decades, yet many questions remain open, in particular the exact geometry of the accretion flow.

Since late 2021, the Imaging X-ray Polarimetry Explorer \citep[IXPE;][]{Weisskopf2022} has been measuring the polarimetric properties of different X-ray sources, bringing remarkable insights into the geometry of WMNS LMXBs in different spectral states. With around 20 of these sources observed so far, several key trends have now been detected \citep[for a review, see][]{Ursini2024}. Among these, IXPE has already observed almost the entire sample of currently known Z-sources. The polarization degree (PD) is consistently higher in the harder state (i.e., the HB) than in the softer states (i.e., the NB and FB) for sources where both states have been observed, i.e., XTE~J1701$-$462 \citep{Cocchi2023}, GX~5$-$1 \citep{Fabiani2024}, and GX~340+0 \citep{LaMonaca2024b}. A general trend of increasing PD with energy is also observed. The spectra of WMNS-LMXBs are usually described as the combination of a soft thermal component and harder Comptonized emission, with some sources also showing a component due to disk reflection \citep{Ludlam2024}. The soft emission seems to exhibit lower polarization, while the polarization of the Compton and reflection components are stronger \citep{Farinelli2023,Cocchi2023,Fabiani2024,LaMonaca2024}. However, the physical origin of the polarization signal and particularly the unexpectedly high PD seen in some sources near the high energy end (8~keV) of the IXPE bandpass is not well understood.

\mbox{Cyg~X-2} is one of the brightest WMNS-LMXBs. It was first observed in the X-rays by \citet{Byram1966} and classified as a binary system by \citet{Sofia1968}. The system has a bright optical counterpart with a mass of $\approx 0.5~M_{\odot}$ \citep{Podsiadlowski2000}, and the mass of the NS is estimated at $\approx 1.7~M_{\odot}$ \citep{Casares2010}. The estimated distance of the source from optical observations is 7.2 $\pm$ 1.1 kpc \citep{Orosz1999}. The source is known to exhibit Type I X-ray bursts clearly indicating an NS accretor \citep[see, e.g.,][]{Kahn1984, Smale1998}. A relativistic jet has been detected while the source was on the HB \citep{Spencer2013}. The jet position angle was $141\degr$.

The X-ray spectrum of Cyg X-2 is usually described as a combination of a soft multi-temperature disk blackbody plus Comptonized emission \citep{DiSalvo2002,Farinelli2009}, characterized by low electron temperature ($kT_{\rm e} \sim 3$ keV) and high Thomson optical depth ($\tau \sim 4-10$, depending on the geometry of the Comptonizing region). In addition, relativistic iron lines and reflection features have been observed in the X-ray spectra \citep{Smale1993,Cackett2010,Ludlam2022}. A recent spectral analysis of the reflection along the Z-track suggests a system inclination between $60\degr$ and $70\degr$ and a rather stable inner disk radius, close to the innermost stable circular orbit \citep{Ludlam2022}.

The first X-ray polarimetric observations of \mbox{Cyg~X-2} were performed with the OSO-8 satellite in 1975--1977 \citep{Long1980}. Polarization was marginally detected in the 1975 data with a polarization degree (PD) of $4.9\% \pm 1.8\%$ and a polarization angle (PA) of $138\degr \pm 10\degr$ at 2.6~keV, but only upper limits on polarization were obtained from the 1976 and 1977 data. IXPE observed \mbox{Cyg~X-2} for the first time in 2022 when the source was moving along the NB and FB through the SA \citep{Farinelli2023,Gnarini2025}. IXPE measured significant X-ray polarization with $\rm{PD}=1.8\% \pm 0.3\%$ and PA=$140\degr \pm 4 \degr$ in the 2--8~keV band, consistent with the radio jet direction and previous OSO-8 measurements. 

Here, we report a new spectropolarimetric observation of \mbox{Cyg~X-2} performed with IXPE and NuSTAR between 2025 May 28 and May 30. The paper is organized as follows. In Section \ref{sec:Observations}, we describe the IXPE and NuSTAR observation with the data reduction procedure. In Section \ref{sec:Polarization}, we report on the model-independent polarimetric analysis using the new \texttt{ixpe\_protractor} tool. In Section \ref{sec:Spectra}, we present the results obtained from spectral and spectropolarimetric analysis. Finally, in Section \ref{sec:Discussion}, we discuss the results obtained, in comparison with previous IXPE observations of Cyg X-2 as well as other Z-sources.

\section{Observations and Data Analysis}\label{sec:Observations}

\subsection{IXPE}

IXPE \citep{Weisskopf2022} is a joint NASA and Italian Space Agency small explorer-class mission capable of measuring the $I$, $Q$, and $U$ Stokes parameters of X-rays in the 2--8~keV band with its three detector units (DUs), each hosting a gas-pixel detector \citep{Baldini2021,Soffitta2021}. IXPE observed \mbox{Cyg X-2} on 2025 May 28 17:03:43 UT to May 30 17:49:49 UT (ObsID: 04250501) for a total exposure time of about 93 ks. Since 2025 April 14, the DU2 has experienced an anomaly which has altered the detector response:\footnote{\url{https://heasarc.gsfc.nasa.gov/docs/ixpe/whatsnew.html}} while the spectral response has already been corrected, the recalibration for the polarimetric response and the background filtering are ongoing. As a result of this issue, for this observation, DU2 data have not been provided and only event data from DU1 and DU3 have been distributed and used during our analysis. 

We extracted the spectral data using the new \texttt{ixpestartx} task\footnote{\url{https://heasarc.gsfc.nasa.gov/docs/ixpe/analysis/contributed/ixpestartx.html}} which uses the standard \textsc{ftools} of HEASoft (v6.35.2; \citealt{HEASoft}) along with the latest calibration files (CALDB v.20250225). Our analysis used the weighted analysis scheme of \cite{DiMarco2022}. We adopted a circular region of 120\arcsec\ radius centered on the source (R.A. 21:44:42.5; Decl. +38:19:04.4) to produce the spectra and light curves. Radii in the range from 30\arcsec\ to 180\arcsec\ with 5\arcsec\ steps were considered and the radius was chosen to maximize the signal-to-noise ratio (S/N) in the 2--8 keV energy band. Since the source is very bright ($> 2$ count~s$^{-1}$~arcmin$^{-2}$), we applied neither background subtraction nor rejection \citep{DiMarco2023}. For each DU, the ancillary response file (ARF) and the modulation response file (MRF) were generated via the \texttt{ixpestartx} script which uses the \texttt{ixpecalcarf} tool, considering the same extraction radius used for the source region. Each $I$ spectrum was rebinned using the \texttt{ftgrouppha} command and requiring a S/N of 5 for each bin, while the $Q$ and $U$ spectra were rebinned using the same energy bins as used for the $I$ spectra. IXPE light curves were extracted using the \texttt{extractor} task. 

\begin{figure}
\centering
\includegraphics[width=0.95\linewidth]{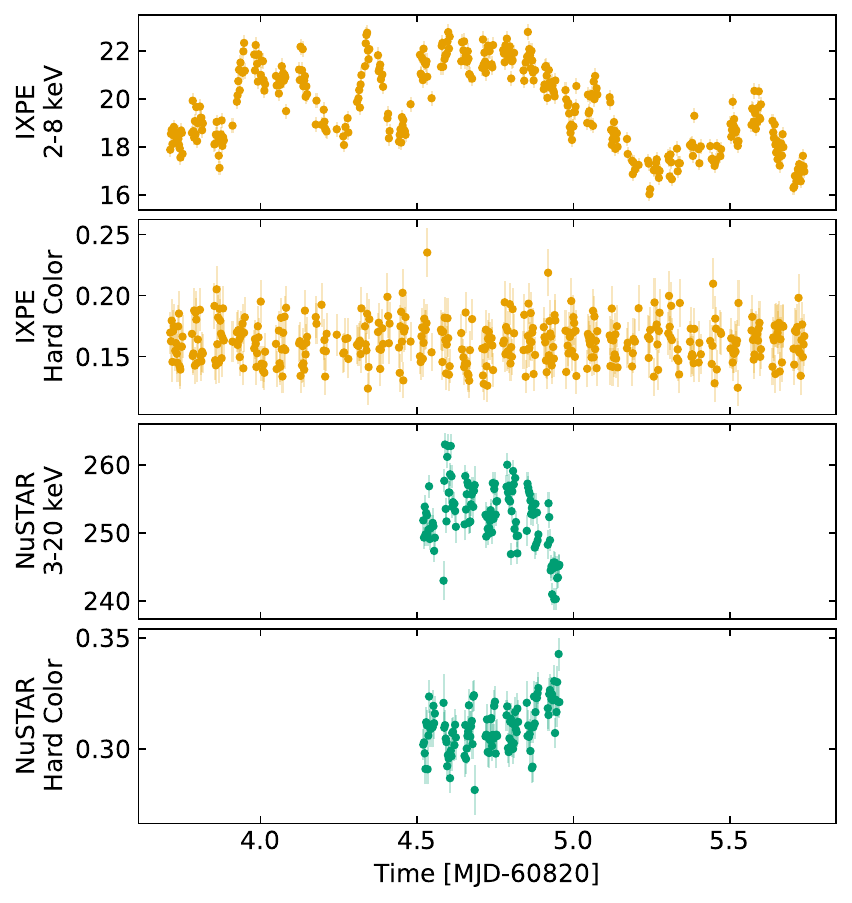}
\caption{IXPE and NuSTAR light curves and hardness ratios of \mbox{Cyg X-2}. The first and third panels show the IXPE (2--8 keV) and \textbf{the background subtracted} NuSTAR (3--20 keV) light curves in count~s$^{-1}$. The second and fourth panels show the IXPE (5--8 keV/3--5 keV) and the NuSTAR (10--20 keV/6--10 keV) hard colors. Each IXPE and NuSTAR time bin corresponds to 200~s.}
\label{fig:LC}
\end{figure}

The total light curve obtained by summing the two DUs with time bins of 200~s is shown in Fig.~\ref{fig:LC}, along with the hard color defined as the ratio of the measured counts in the 5--8 keV/3--5 keV energy bands. Although the hard color remains roughly constant throughout the observations, the source shows variability in its 2--8 keV emission.

\begin{figure}
\centering
\includegraphics[width=0.95\linewidth]{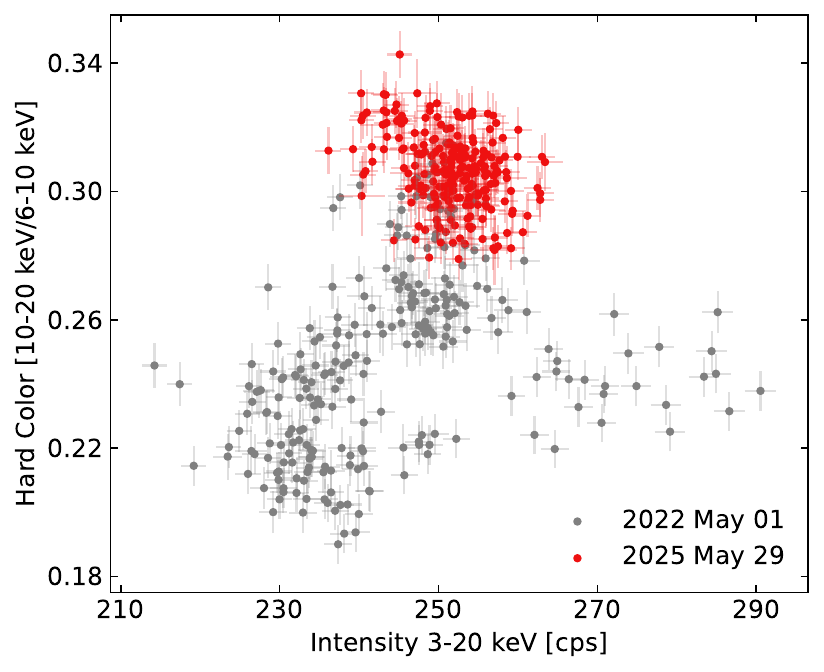}
\caption{NuSTAR hardness-intensity diagram of \mbox{Cyg X-2} from observations simultaneous with IXPE. The intensity (in count~s$^{-1}$) is considered in the 3--20 keV range, while the hard color is defined as the ratio of the counts in the 10--20 keV/6--10 keV bands, after background subtraction. Gray points represent the 2022 observation, while the red ones highlight the new NuSTAR observation in 2025. Each point corresponds to 200~s.}
\label{fig:NuSTAR.HID}
\end{figure}

\subsection{NuSTAR}

The Nuclear Spectroscopic Telescope Array (NuSTAR; \citealt{Harrison2013}) observed \mbox{Cyg X-2} on 2025 May 29 12:28:55 UT to 22:58:52 UT (ObsID: 11001313004) for a net exposure time of about 14~ks for each focal plane module (FPM) telescope. Data were processed using the standard \texttt{nupipeline} task of NuSTAR Data Analysis Software (\textsc{nustardas} v.2.1.5) with the latest calibration files (CALDB v.20250317). Due to the high brightness of \mbox{Cyg X-2} ($> 100$ count~s$^{-1}$), during the data processing with \texttt{nupipeline}, we filtered the data using the \texttt{statusexpr="(STATUS==b0000xxx00xxxx000)\& \&(SHIELD==0)"} keyword. 

Similarly to IXPE, we considered circular regions centered on the source (R.A. 21:44:40.1; Decl. +38:19:17.4) with radii computed using the same procedure to maximize the S/N (see also \citealt{Piconcelli2004}). In this case, we obtained a source region radius of 180\arcsec. We performed background subtraction using a circular region with 60\arcsec\ radius in an area of the detector with negligible source counts. The source and background spectra, along with the light curves, were obtained using the \texttt{nuproducts} task. Then, we regrouped the spectra using \texttt{ftgrouppha}, considering the optimal binning algorithm by \cite{Kaastra2016}, with a minimum S/N of 3 per grouped bin. 

To characterize the state of the source during this observation, we generated the NuSTAR HID (Figure~\ref{fig:NuSTAR.HID}) considering the 3--20 keV intensity and the hard color defined as the ratio of the measured counts in the 10--20/6--10 keV energy range, after background subtraction. \mbox{Cyg X-2} exhibited harder emission in 2025 than in 2022. The source was clearly in the HB of its Z-track during the 2025 NuSTAR observation presented in this work. The IXPE hard color remains constant throughout the observation while the flux varies by only $\pm 16\%$, see Fig.~\ref{fig:LC}. Thus, \mbox{Cyg X-2} remains in the same spectral state. We conclude that the source remained in the HB throughout the IXPE observation. Therefore, for the spectral and polarimetric analysis, we considered the average spectra throughout the entire observation. This is the first IXPE observation of Cyg X-2 in the HB.

\begin{deluxetable}{ccc}
\tablecaption{X-ray polarization \textbf{of Cyg X--2} for different energy bands obtained with the \texttt{ixpepolarization} task.\label{tab:Polarization}}
\tablehead{\colhead{Energy Bin} & \colhead{PD} & \colhead{PA} \\
\colhead{(keV)} & \colhead{(\%)} & \colhead{(deg)}} 
\startdata
2--8  & 4.5 $\pm$ 0.3 & 128 $\pm$ 2 \\
\hline
2--4  & 3.8 $\pm$ 0.4 & 128 $\pm$ 3 \\
4--6  & 4.7 $\pm$ 0.6 & 128 $\pm$ 4 \\
6--8  & 8.1 $\pm$ 1.4 & 134 $\pm$ 5 \\
\hline
2--3  & 3.3 $\pm$ 0.5 & 130 $\pm$ 4 \\
3--4  & 3.9 $\pm$ 0.5 & 126 $\pm$ 4 \\
4--5  & 4.3 $\pm$ 0.7 & 129 $\pm$ 5 \\
5--6  & 5.3 $\pm$ 1.0 & 127 $\pm$ 6 \\
6--7  & 7.5 $\pm$ 1.6 & 134 $\pm$ 6 \\
7--8  & 9.9 $\pm$ 2.8 & 132 $\pm$ 8 \\
\enddata
\tablecomments{Errors correspond to the 68\% confidence level.}
\end{deluxetable}

\section{Polarimetric Analysis}\label{sec:Polarization}

We first examined the X-ray polarization of \mbox{Cyg X-2} in a model-independent manner using the \texttt{ixpepolarization} task of HEASoft included in the new \texttt{ixpe\_protractor} tool.\footnote{\url{https://heasarc.gsfc.nasa.gov/docs/ixpe/analysis/contributed.html}} The \texttt{ixpepolarization} task computes the overall Stokes parameters for a selected spatial region and energy range. The \texttt{ixpe\_protractor} tool reads the output Stokes parameters and generates a ``protractor'' plot of polarization contours in PD and PA.

\begin{figure}
\centering
\includegraphics[width=0.9\linewidth]{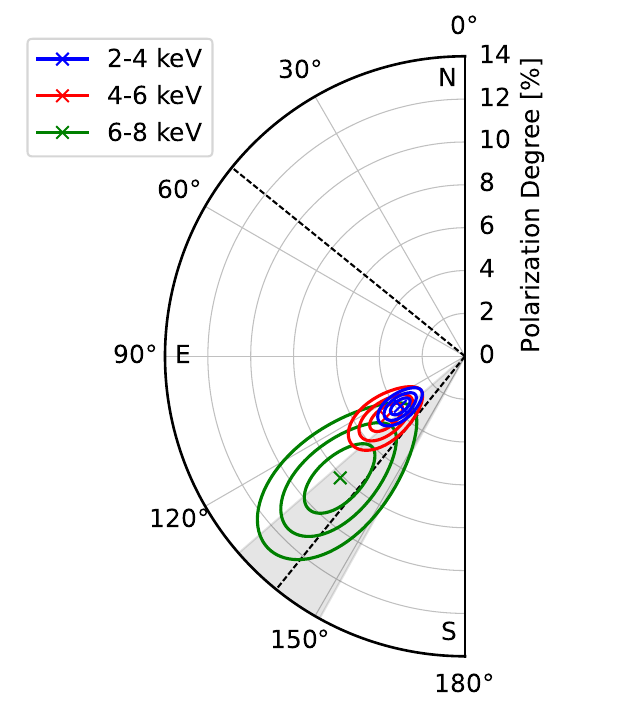}
\caption{Polarization contours in the 2--4 keV, 4--6 keV and 6--8 keV band at the 68\%, 90\% and 99\% confidence levels obtained with the \texttt{ixpe\_protractor} task. The gray region highlights the direction of the radio jet \citep{Spencer2013}, \textbf{and the black, dashed lines indicate the radio jet direction and the orthogonal direction}.}
\label{fig:Protractor-En}
\end{figure}

\begin{figure}
\centering
\includegraphics[width=0.95\linewidth]{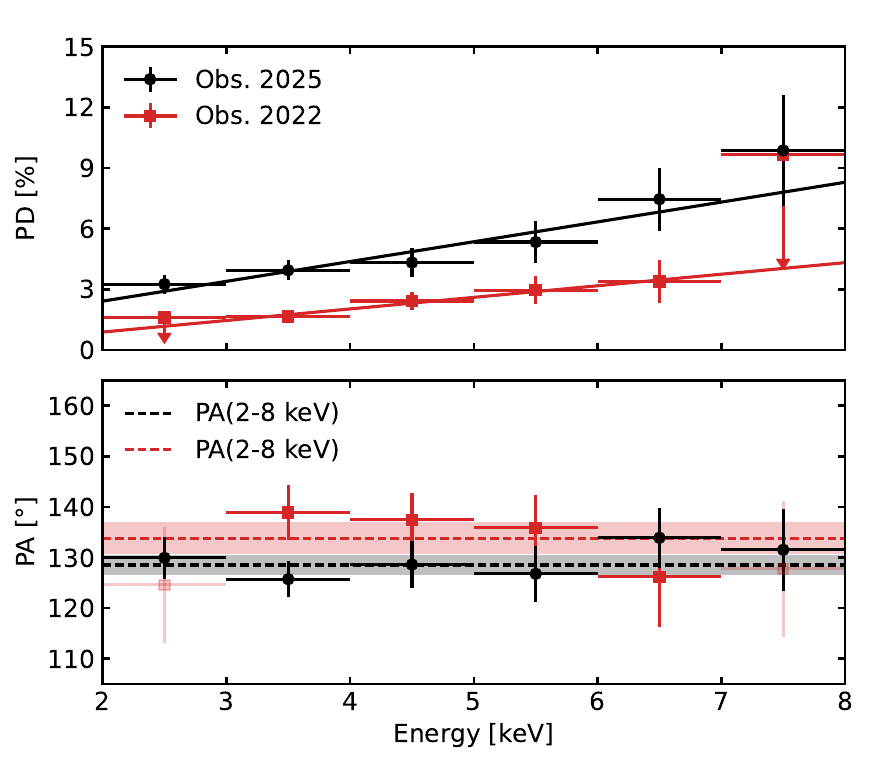}
\caption{Polarization degree (top) and angle (bottom) versus energy in 1~keV bins. The black points correspond to 2025 May 29 observation (i.e., HB) and the red points are the sum of 2022 April 30 and 2022 May 2 observations (i.e., NB and FB; \citealt{Gnarini2025}). Solid lines in the upper panel represent the best linear fits. Dashed lines in the lower panel correspond to the average PA in the 2--8 keV range, with the associated $1\sigma$ error highlighted by the shaded regions.}
\label{fig:PD_PA_Energy}
\end{figure}

With this model-independent procedure, we detected significant polarization in the 2--8 keV range at 15$\sigma$ significance with PD = 4.5\% $\pm$ 0.3\% and PA = $128\degr \pm 2\degr$.\footnote{All the uncertainties in the text related to our polarization measurements are reported at the $1\sigma$ confidence level.} The polarization contours computed using \texttt{ixpe\_protractor} considering 2 keV-energy bin are reported in Figure~\ref{fig:Protractor-En}: for each bin, the polarization is well constrained at more than 99\% confidence level. Use of narrower energy intervals (e.g., 1~keV-; Table~\ref{fig:PD_PA_Energy}) reveals a significant energy dependence. The PD increases with energy, up to $9.9\%\pm 2.8\%$ in the 7--8 keV band, with no indication of a rotation in the PA. To properly assess the significance of the increasing trend of the PD with energy, we performed a linear fit to the results obtained with \texttt{ixpe\_protractor}: the linear model provides a statistically significant better fit to the data with respect to a fit with constant PD ($p$-value = 0.004), corresponding to a 99.6\% confidence level. The linear fits are shown in Fig.~\ref{fig:PD_PA_Energy}. All polarimetric results derived with \texttt{ixpe\_protractor} are consistent with those derived using the \texttt{PCUBE} task of \textsc{ixpeobssim} \citep{Baldini2022}. 

We also tried to study the dependence of the polarization on the source flux by separating the time intervals with higher counts ($\gtrsim 20$ count~s$^{-1}$) from those with lower ($\lesssim 20$ count~s$^{-1}$). The resulting polarization for the different flux levels is compatible within errors: a PD of 4.3\% $\pm$ 0.5\% with PA = $130\degr \pm 4\degr$ is obtained for the high-flux intervals, while we found a PD of 4.6\% $\pm$ 0.4\% with PA = $128\degr \pm 4\degr$ for interval with lower flux. 

Figure~\ref{fig:PD_PA_Energy} compares the energy dependence of the PD in the current observation with that in the previous ones using 1~keV-energy bins. In particular, following \cite{Gnarini2025}, we combined the results obtained with the \texttt{ixpe\_protractor} task for two IXPE observations performed between 2022 April and May, since the state of the source is the same and the Stokes parameters are consistent between the two observations (see also Figures 3 and 4 in \citealt{Gnarini2025}). The most significant difference between the 2022 observations and the new one are the higher PD values in 2025 (Fig.~\ref{fig:PD_PA_Energy}), as seen already for the average PD. Both observations show increasing PD with energy, without indication of rotation in the PA. The PA is roughly consistent with that measured for \mbox{Cyg X-2} in the NB and FB \citep{Farinelli2023,Gnarini2025} and with the position angle of a discrete radio jet ejection \citep{Spencer2013}.

\section{Spectropolarimetric Analysis}\label{sec:Spectra}

We first fit the joint IXPE (2--8 keV) and NuSTAR (3--30 keV) spectra using \textsc{xspec}. Since the source remains in the HB, we considered the average spectra throughout the observation. We used the following spectral model:
\begin{displaymath}
{\tt tbabs*(diskbb+thcomp*bbodyrad+relxillNS)}.
\end{displaymath}
We model interstellar absorption with \texttt{tbabs} and measure the hydrogen column density $N_{\rm H}$ adopting \texttt{vern} cross-section \citep{Verner1996} and \texttt{wilm} abundances \citep{Wilms2000}. The baseline continuum consists of two components: a multicolor disk blackbody (\texttt{diskbb}; \citealt{Mitsuda1984}) and a harder Comptonized blackbody (\texttt{thcomp*bbodyrad}; \citealt{Zdziarski2020}). We left free to vary during the fit all the physical parameters of these two components. However, the fraction of Comptonized seed photons represented by the covering factor $f$ of \texttt{thcomp} turns out to be $>0.99$ and we decided to fix this parameter to its best-fit value ($f = 1$) in order to find better constraints for the other parameters of \texttt{thcomp*bbodyrad}. 

\begin{figure}
\centering
\includegraphics[width=0.95\linewidth]{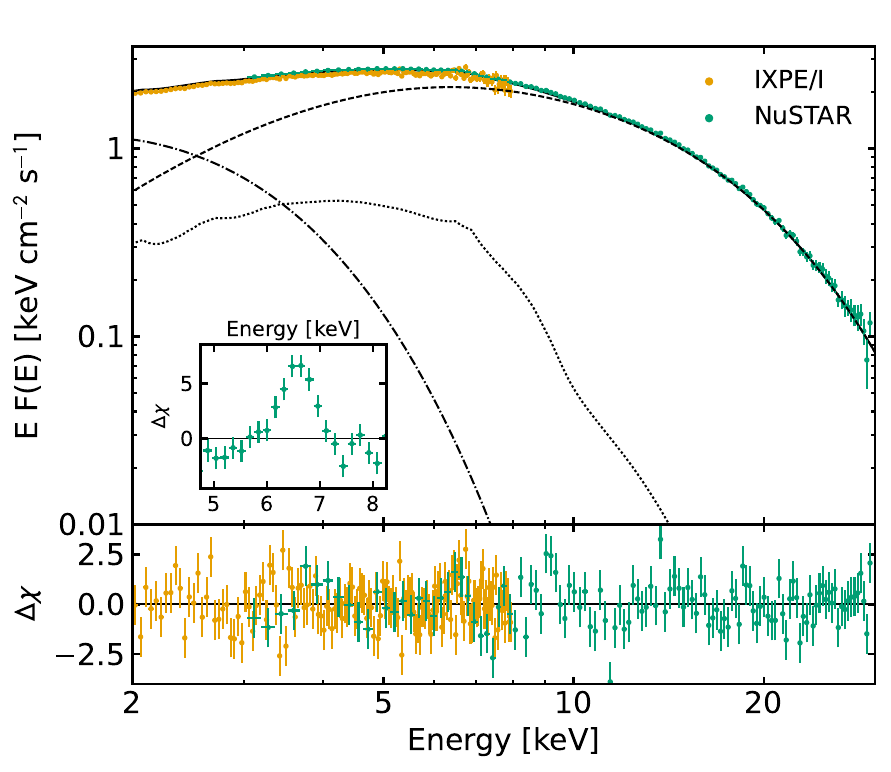}
\caption{IXPE (2--8 keV) and NuSTAR (3--30 keV) deconvolved spectra with the resulting best-fit model and the corresponding residuals in units of $\sigma$.  The model is composed of \texttt{diskbb} (dash-dotted lines), \texttt{thcomp*bbodyrad} (dashed lines), and \texttt{relxillNS} (dotted lines) components. The inset shows the residuals without the reflection component, highlighting the Fe K$\alpha$ line profile.}
\label{fig:Spectra}
\end{figure}

\begin{deluxetable}{ll c}
\tablewidth{\linewidth}
\tablecaption{Best-fit spectral model parameters.\label{tab:Best-Fit}}
\tablewidth{\linewidth}
\tablehead{ & Parameter & Value} 
\startdata
\texttt{tbabs} & $N_{\rm H}$ ($10^{22}$\,cm$^{-2}$) & [0.12] \\
\texttt{diskbb} & $kT_{\rm in}$ (keV) & 0.65$^{+0.02}_{-0.02}$ \\
 & $R_{\rm in} \sqrt{\cos i}$ (km) & 23.3$^{+0.8}_{-0.8}$ \\
\texttt{thcomp} & $kT_{\rm e}$ (keV) & 3.4$^{+0.1}_{-0.1}$ \\ 
 & $\tau$ & 8.5$^{+0.4}_{-0.5}$ \\
 & $f$ & [1] \\
\texttt{bbodyrad} & $kT$ (keV) & 1.10$^{+0.01}_{-0.01}$ \\
 & $R_{\rm bb}$ (km) & 11.6$^{+0.3}_{-0.3}$ \\ 
\texttt{relxillNS} & $q_{\rm em}$ & 1.6$^{+0.2}_{-0.2}$ \\ 
 & $a$ & [0.1] \\
 & $i$ (deg) & [62] \\ 
 & $R_{\rm in}$ (ISCO) & $<5.3$ \\ 
 & $kT_{\rm bb}$ (keV) & = $kT$ \\ 
 & $\log \xi$ & 2.96$^{+0.06}_{-0.07}$ \\ 
 & $A_{\rm Fe}$ & [1.4] \\ 
 & $\log N_{\rm e}$/cm$^{-3}$ & [18] \\ 
 & $N_{\rm r}$ ($10^{-3}$) & 2.8$^{+0.9}_{-0.8}$ \\
\hline
\multicolumn{3}{c}{Cross-calibration} \\
& $\mathcal{C}_{\rm DU3/DU1}$ & 0.970$^{+0.003}_{-0.002}$ \\
& $\mathcal{C}_{\rm FPMA/DU1}$ & 1.372$^{+0.004}_{-0.004}$ \\
& $\mathcal{C}_{\rm FPMB/DU1}$ & 1.392$^{+0.005}_{-0.004}$ \\
\hline 
\multicolumn{2}{l}{$\chi^2/{\rm d.o.f.}$} & 549/526 \\
\hline
\multicolumn{3}{c}{Photon flux ratios (2--8 keV)} \\
\multicolumn{2}{l}{$N_{\tt diskbb}/N_{\rm Tot}$} & 24.6\% \\
\multicolumn{2}{l}{$N_{\tt thcomp*bbodyrad}/N_{\rm Tot}$} & 57.6\% \\
\multicolumn{2}{l}{$N_{\tt relxillNS}/N_{\rm Tot}$} & 17.8\% \\
\multicolumn{3}{c}{Photon flux ratios (2--4 keV)} \\
\multicolumn{2}{l}{$N_{\tt diskbb}/N_{\rm Tot}$} & 35.7\% \\
\multicolumn{2}{l}{$N_{\tt thcomp*bbodyrad}/N_{\rm Tot}$} & 46.3\% \\
\multicolumn{2}{l}{$N_{\tt relxillNS}/N_{\rm Tot}$} & 18.0\% \\
\multicolumn{3}{c}{Photon flux ratios (4--6 keV)} \\
\multicolumn{2}{l}{$N_{\tt diskbb}/N_{\rm Tot}$} & 6.5\% \\
\multicolumn{2}{l}{$N_{\tt thcomp*bbodyrad}/N_{\rm Tot}$} & 74.5\% \\
\multicolumn{2}{l}{$N_{\tt relxillNS}/N_{\rm Tot}$} & 19.0\% \\
\multicolumn{3}{c}{Photon flux ratios (6--8 keV)} \\
\multicolumn{2}{l}{$N_{\tt diskbb}/N_{\rm Tot}$} & 0.9\% \\
\multicolumn{2}{l}{$N_{\tt thcomp*bbodyrad}/N_{\rm Tot}$} & 85.3\% \\
\multicolumn{2}{l}{$N_{\tt relxillNS}/N_{\rm Tot}$} & 13.8\% \\
\multicolumn{3}{c}{Energy Flux (2--8 keV)} \\
\multicolumn{2}{l}{$F_{\rm Tot}$ ($10^{-9}$ \fluxcgs)} & $5.38 \pm 0.05$ \\
\enddata
\tablecomments{Errors are at the 90$\%$ confidence level for a single parameter. The parameters in square brackets were kept frozen during the fit. The radii of \texttt{diskbb} and \texttt{bbodyrad} are calculated assuming a source distance of 7.2 kpc \citep{Orosz1999}.}
\end{deluxetable}

\begin{figure*}
\centering
\includegraphics[width=0.45\linewidth]{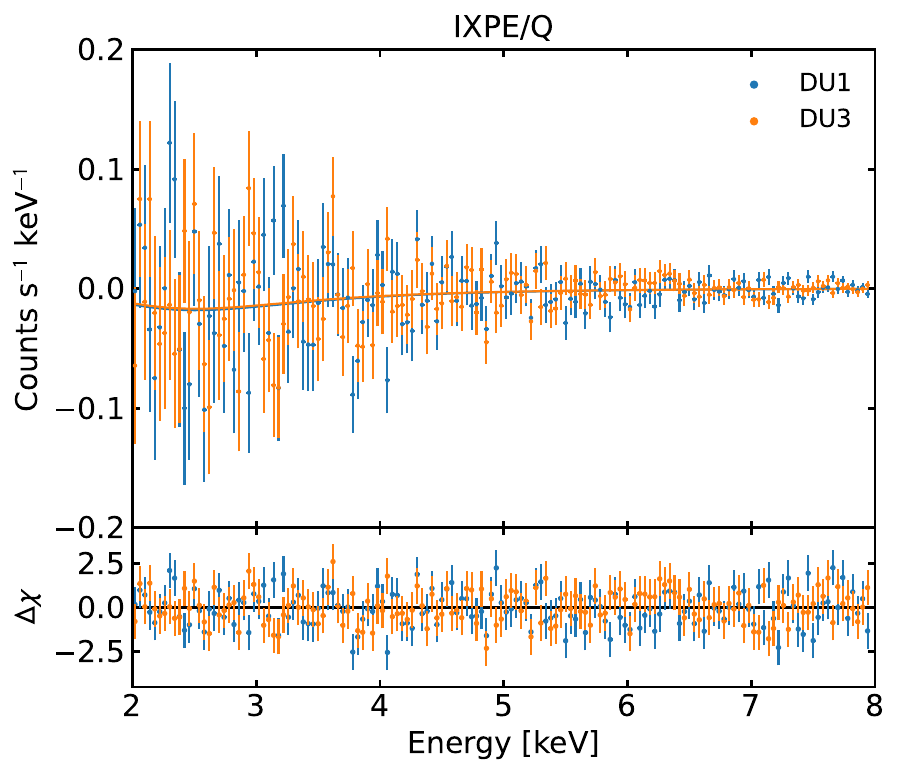}
\includegraphics[width=0.45\linewidth]{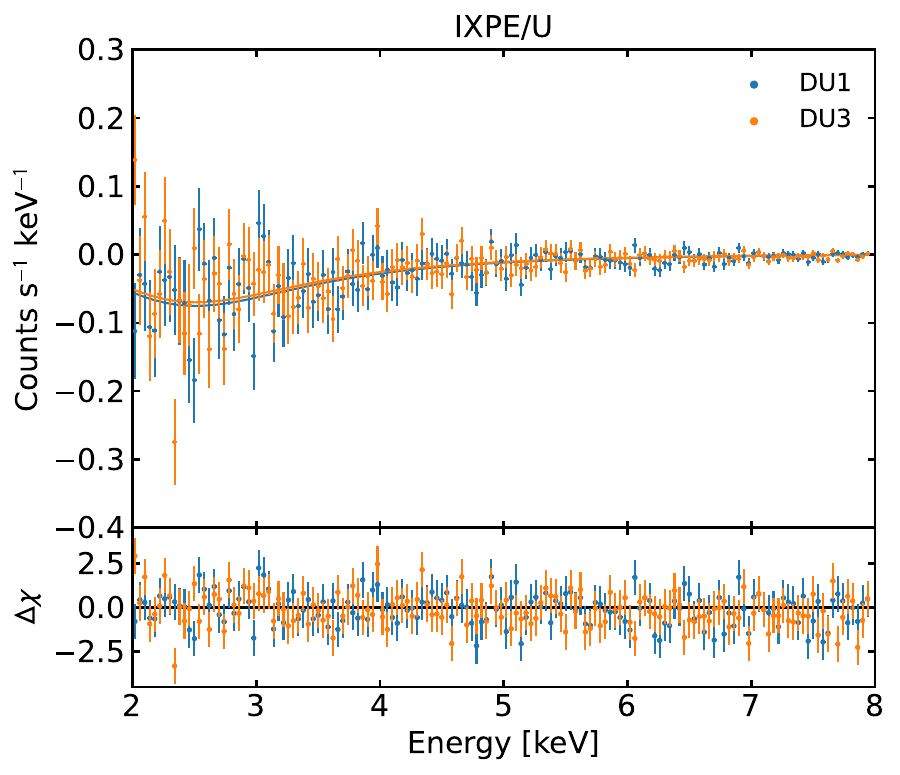}
\caption{IXPE $Q$ and $U$ Stokes spectra for each DU and the corresponding residuals in units of $\sigma$, obtained applying \texttt{polconst} to best-fitting spectral model (Table \ref{tab:Best-Fit}).}
\label{fig:Stokes.Spectra}
\end{figure*}

The NuSTAR residuals for this baseline model clearly show the presence of a strong iron line (inset plots in Figure \ref{fig:Spectra}). Therefore, we included a \texttt{relxillNS} \citep{Garcia2022} component to model photons reflected off the accretion disk. This model assumes a blackbody spectrum that irradiates the surface of the accretion disk incident at $45\degr$. We assumed that the the temperature of the seed photons of \texttt{relxillNS} $kT_{\rm bb}$ is tied to the blackbody temperature of \texttt{bbodyrad}. We chose to fix certain parameters that the fit fails to constrain or does not yield reasonable values: the dimensionless spin $a$ is fixed at 0.1, a typical value for WMNS-LMXBs \citep{DiSalvo2024}; the number density is set to $\log N_{\rm e}$/cm$^{-3}$ = 18, consistent with the results found by \cite{Ludlam2022} and with typical values for standard accretion disks \citep{Garcia2016}; the iron abundance $A_{\rm Fe}$ is fixed at 1.4, as derived by \cite{Ludlam2022}; and the outer disk radius is set at $R_{\rm out} = 1000$ $R_{\rm g}$. Conversely, the emissivity index $q_{\rm em}$, the ionization parameter $\xi$, and the inner disk radius $R_{\rm in}$ were left free to vary during the fit. We decided to fix also the inclination to the value of $62\degr$ obtained by \citet{Ludlam2022}.

Figure \ref{fig:Spectra} shows the IXPE and NuSTAR deconvolved spectra with the resulting best-fit spectral model. The best-fit parameters are reported in Table \ref{tab:Best-Fit}. The addition of \texttt{relxillNS} provides a better fit, with a $\chi^2$/d.o.f. of 549.5/526. The physical parameters of the disk and Comptonized components are constrained with good precision and align with typical values for Cyg X-2 \citep[e.g.,][]{DiSalvo2002,Farinelli2009}. We computed the radius of the black body photon-emitting region $R_{\rm bb}$ and the apparent inner disk radius $R_{\rm in}$ from the normalizations of \texttt{bbodyrad} and \texttt{diskbb}, respectively, assuming a source distance of 7.2 kpc \citep{Orosz1999}. We did not detect the transient hard tail, which was previously observed in the spectra of \mbox{Cyg X-2} during the HB with INTEGRAL and BeppoSAX \citep{Paizis2006,Farinelli2009}, but we note that the NuSTAR background is dominant for energies above 30~keV. The ionization parameter and the emissivity index for the reflection component are well constrained, but only an upper limit is found for the inner disk radius; this limit is not very stringent and it is consistent with the results of \cite{Ludlam2022}.

Once the best-fit model to the spectra was determined, we performed spectropolarimetric fits with all spectral parameters fixed to their best-fit values and using only the IXPE $I$, $Q$, and $U$ spectra (Table~\ref{tab:Best-Fit}). We first applied \texttt{polconst} to the whole spectral model. The \texttt{polconst} model assumes that the PD and the PA are constant with energy. The IXPE $Q$ and $U$ Stokes spectra and the fits with this model are shown in Figure~\ref{fig:Stokes.Spectra}. The polarimetric results are consistent with those derived with the \texttt{ixpe\_protractor} task for each energy bin. We also compared the \texttt{polconst} model with the linear energy-dependent model \texttt{pollin} to test the PD variation with energy. Since no significant rotation in the PA with energy is observed, it is assumed to be constant, while the PD varies with energy $E$ as:
\begin{equation}
    {\rm PD}(E) = {\rm PD_{2keV}} + (E - 2~{\rm keV}) \times A_{\rm PD}~,
\end{equation}
where ${\rm PD_{2keV}}$ is the PD at 2~keV and $A_{\rm PD}$ is the slope of the linear dependence with energy. The fit improves with respect to the \texttt{polconst} case with a reduced $\chi^2$ ($\chi^2/\mathrm{dof} = 834/891$ for \texttt{polconst} and $827/890$ for \texttt{pollin}). The F-test yields a $p$-value of 0.004, corresponding to a 99.6\% confidence level. The energy dependence is consistent with the results obtained with the \texttt{ixpe\_protractor} tool. The results obtained from \texttt{pollin} indicate that as the energy increases, the PD also increases from a value of $3.0\% \pm 0.8\%$ at 2 keV with a slope of $(0.8 \pm 0.3)\%\,{\rm keV}^{-1}$.

We then applied \texttt{polconst} or \texttt{pollin} separately to the three spectral component (i.e., \texttt{diskbb}, \texttt{thcomp*bbodyrad} and \texttt{relxillNS}). Leaving the PD and PA for each spectral component free to vary produced no useful constraints on the polarization. Therefore, we considered two cases with different assumptions about the polarization of each spectral component. Specifically, the PD of the \texttt{thcomp*bbodyrad} component is constant with energy in Case~1 versus a linear dependence of PD for \texttt{thcomp*bbodyrad} in Case~2. In both cases, the PD of the disk emission is constant with energy, and the PD of the reflected photons was fixed at 10\% \citep{Matt1993}. We assumed that the PA of \texttt{relxillNS} and \texttt{thcomp*bbodyrad} are the same \citep[see, e.g.,][]{Farinelli2023,Ursini2023}. This assumption is consistent with a geometry in which the optically thick BL/SL is characterized by a vertical height significantly greater than its radial extension ($H \gg \Delta R$) \citep{Matt1993,Poutanen1996,Schnittman2009}. 

\begin{deluxetable}{l ccc}
\tablecaption{Polarization degree and angle of each spectral component.\label{tab:Pol-Fit}}
\tablehead{\colhead{Component} & \colhead{PD$_{\rm 2keV}$} & \colhead{$A_{\rm PD}$} & \colhead{PA} \\
& \colhead{(\%)} & \colhead{(\%\,keV$^{-1}$)} & \colhead{(deg)}}
\startdata
\multicolumn{4}{c}{\textbf{Case 1}} \\
\texttt{diskbb} & $<2.6$ & [0] & = PA$_{\tt thcomp} - 90$ \\
\texttt{thcomp} & $4.2 \pm 0.9$ & [0] & $128 \pm 3$ \\
\texttt{relxillNS} & [10] & [0] & = PA$_{\tt thcomp}$ \\
\multicolumn{4}{c}{$\chi^2/{\rm d.o.f.} = 818.7/889$} \\
\hline
\multicolumn{4}{c}{\textbf{Case 2}} \\
\texttt{diskbb} & $<2.3$ & [0] & = PA$_{\tt thcomp} - 90$ \\
\texttt{thcomp} & $2.9 \pm 0.9$ & $0.7 \pm 0.3$ & $128 \pm 4$ \\
\texttt{relxillNS} & [10] & [0] & = PA$_{\tt thcomp}$ \\
\multicolumn{4}{c}{$\chi^2/{\rm d.o.f.} = 814.8/888$} \\
\enddata
\tablecomments{Case~1 has constant PD with energy for the \texttt{thcomp*bbodyrad} component. Case~2 has a linear linear dependence of PD on energy for the \texttt{thcomp*bbodyrad} component. Errors are at the 90$\%$ confidence level for a single parameter. The parameters in square brackets were kept frozen during the fit.}
\end{deluxetable}

The results are reported in Table \ref{tab:Pol-Fit}. For both cases, the $\chi^2$/d.o.f. are significantly better than when applying \texttt{polconst} or \texttt{pollin} to the whole model, so they better reproduce the change of PD with energy. For Case 1, we obtained a PD of $4.2\% \pm 0.9\%$ for Comptonized photons, which are the main contribution to the polarization. We obtained only an upper limit of 2.6\% for disk photons, consistent with the classical value for a semi-infinite, plane-parallel electron-scattering dominated atmosphere seen at an inclination of about $60\degr$ \citep{Chandrasekhar1960,Sobolev1963}. Case 2, with an energy-dependent PD for the Comptonized emission motivated by the significant increase in PD with energy in the model-independent analysis, led to an improved fit. The F-test between the \texttt{polconst} and \texttt{pollin} cases yields a $p$-value of 0.039, indicating that the improvement of the fit is statistically significant at the 96\% confidence level. The best-fitted parameters for \texttt{pollin} indicate that the PD of \texttt{thcomp*bbodyrad} increases with energy, with a PD at 2 keV of $2.9\% \pm 0.9\%$ and a slope of $(0.7 \pm 0.3)~\%\,{\rm keV}^{-1}$, both consistent with the results of the model-independent analysis.

\begin{figure}
\centering
\includegraphics[width=0.95\linewidth]{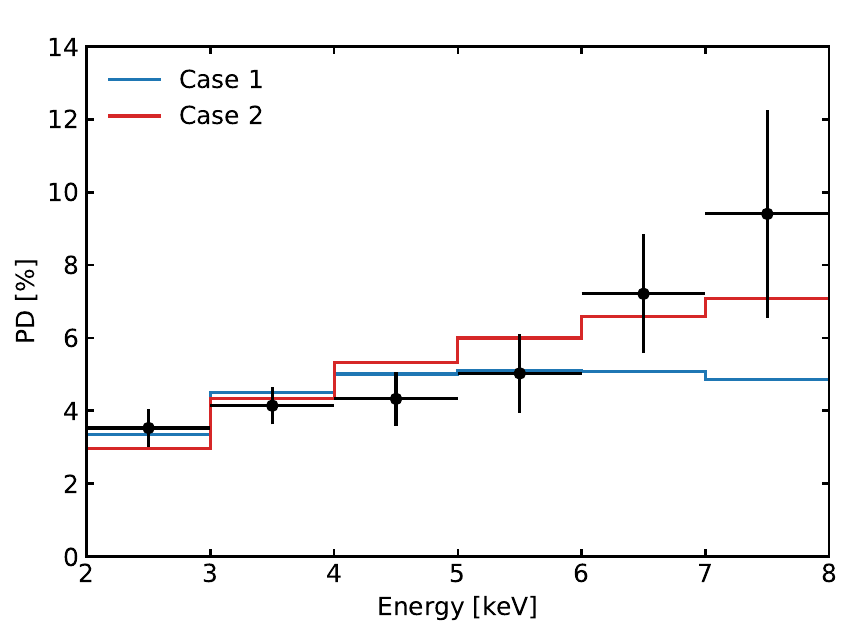}
\caption{Total PD obtained combining the contribution of the three spectral components (i.e., \texttt{diskbb}, \texttt{thcomp*bbodyrad}, and \texttt{relxillNS}) for each case considered in the spectropolarimetric analysis (Table~\ref{tab:Pol-Fit}). Black points correspond to the PD in each bin derived applying \texttt{polconst} to the whole spectral model. Errors correspond to the 90\% confidence level.} 
\label{fig:PD-Energy-Models}
\end{figure}

In Figure \ref{fig:PD-Energy-Models}, we compared the total PD by combining the contribution of the three components. To compute the total polarization, we follow the approach described in \cite{Ursini2023}, summing the Stokes parameters for each component calculated from the values of the PD and PA derived with \textsc{Xspec} (Table \ref{tab:Pol-Fit}) and weighting by the photon flux ratios in each 1~keV energy bin. We note that the case with \texttt{polconst} applied to each components is not able to produce a PD consistent with the measured PD at high energies (above 6~keV; Figure \ref{fig:PD-Energy-Models}), while the case with \texttt{pollin} applied to \texttt{thcomp*bbodyrad} and \texttt{polconst} applied to \texttt{diskbb} and \texttt{relxillNS} better reproduces the data and the trend of increase PD with energy.

\section{Discussion}\label{sec:Discussion}

During the new IXPE observation performed in 2025 May, Cyg X-2 was in the HB, which is the hardest spectral state of Z-sources. The average PD in the 2--8~keV band was $4.5\% \pm 0.3\%$, which is significantly higher than the PD of $1.8\% \pm 0.3\%$ measured during the previous IXPE observation, when \mbox{Cyg X-2} was moving along the NB and FB \citep{Farinelli2023,Gnarini2025}. Our results align with the trends discovered previously in IXPE observations of Z-sources. The average PD in the HB that we observed for \mbox{Cyg X-2} is similar to that measured for other Z-sources with IXPE observations in the HB (XTE~J1701$-$462, \citealt{Cocchi2023}; \mbox{GX 5$-$1}, \citealt{Fabiani2024}; and GX~340+0, \citealt{LaMonaca2024b}), all close to 4\%. The trend of higher PD in the HB as compared with NB is consistent with that found for other Z-sources \citep{Gnarini2025}. 

Our energy-resolved, model-independent analysis showed a significant increasing trend of the PD with energy. In the HB observation presented here, the PD of Cyg X-2 reaches $9.9\% \pm 2.8\%$ in the 7--8~keV band, which is remarkably high. Again, this aligns with the trends seen in other WMNS binaries when sufficient statistics are available \citep{Ursini2024,Gnarini2025}, notably the PD of $12.2\% \pm 3.6\%$ in the 7.5--8.0~keV band seen from the Z source \mbox{GX~340+0} in the hard state \citep{LaMonaca2024b} and the PD of $10\% \pm 2\%$ seen from the atoll source \mbox{4U~1820$-$303} in the 7--8~keV band \citep{DiMarco2023_4U1820}. During our IXPE observation of \mbox{Cyg X-2}, the PA was aligned with the radio jet and showed no evidence for variation with energy.

Our spectroscopic analysis reveals that the spectrum in the IXPE band is dominated by Comptonized emission which contributes $\sim$60\% of the total 2--8~keV flux (see Table \ref{tab:Best-Fit}). The Comptonized flux fraction increases with energy, reaching $\sim$85\% in the 6--8~keV band. The next largest contributor is the disk emission with $\sim25$\% of the 2--8~keV flux. The disk emission drops rapidly with energy to $\sim$7\% in the 4--6~keV band and is negligible above 6~keV. The NuSTAR spectrum indicates the presence of an Fe K$\alpha$ emission line, and a reflection component is required to accurately model the data. Reflection contributes $\sim$18\% of the total 2--8~keV flux. Its contribution is roughly constant up to $\sim$4~keV and drops to $\sim14$\% in the 6--8~keV band.

The high degree of polarization seen from \mbox{Cyg X-2} in the HB, particularly at high energies, challenges theoretical models. The high PD and increase of PD with energy suggest that the Comptonization and/or reflection components are the main contributors to the observed polarization. 

None of the theoretical models for Comptonized BL/SL emission predicts a PD consistent with the results for Cyg X--2 and other Z-sources in the HB in the 2--8 keV range \citep{Gnarini2022,Farinelli2024,Bobrikova2025} and with the PA aligned with the jet. As discussed above, the Comptonized emission may arise from either a boundary layer (BL) or spreading layer (SL). Since the BL is coplanar to the accretion disk, its radiation is expected to be polarized along the disk plane \citep[see, e.g.,][]{Dovciak2008,Loktev2022}, i.e. perpendicular to the jet and orthogonal to the PA observed from \mbox{Cyg X-2}. For a SL on the NS surface, the PA is orthogonal to the disk plane as observed in the case of \mbox{Cyg X-2}, but the PD likely does not exceed 1.5\% \citep[see, e.g.,][]{Gnarini2022,Farinelli2024,Bobrikova2025}. Thus, typical SL geometries do not produce the high polarization observed.

Reflected photons are expected to be polarized at a 20\% level for edge-on observers and at $\sim$10\% for inclinations near $60\arcdeg$ as inferred for \mbox{Cyg X-2} \citep{Matt1993,Poutanen1996}. Thus, it may be possible to produce the observed PD from the combination of a highly polarized reflected component and a moderately polarized SL (see \citealt{Lapidus1985}). This was suggested to explain the high PD observed in \mbox{4U~1820$-$303} \citep{DiMarco2023_4U1820}. This physical situation is represented by Case~1 of our spectropolarimetric modeling (Table \ref{tab:Pol-Fit}). The decreasing disk flux contribution with energy versus the increasing contribution of Comptonized and reflected X-rays provides a reasonable fit to the PD energy dependence below 6~keV. This could also help explain the decreased PD in the soft state, since the disk is then more prominent \citep[see][]{Farinelli2023}. However, the PD of the Comptonized emission is higher than produced by typical SL geometries, and the model PD lies well below that measured above 6~keV. Our Case~2 model provides a better fit, particularly at high energies. However, it requires that the polarization of the Comptonized emission increase with energy. The very high PD in the 7--8~keV band is well beyond that produced in typical SL geometries.

More sophisticated disk models may yield stronger reflection PD and an increasing PD with energy. \citet{Podgorny2025} showed that the degree of polarization depends on the disk geometry, the ionization state of the disk, and the spectral shape of the incident radiation. The results suggest that the observed increase of X-ray polarization with energy in IXPE observations could be related to the intrinsic polarization of reflected thermal emission for highly ionized slabs. This trend arises from the inelastic Compton scatterings, producing an increase of polarization with energy whose slope depends on the incident spectrum and the reflection geometry.

Electron scattering in a wind above the accretion disk could be another way to produce such a high polarization (parallel to the disk axis). This was considered as a potential explanation for the high PD seen in \mbox{GX~5$-$1} \citep{Fabiani2024}. In the wind scenario, higher energy photons have decreased absorption in the wind and the higher scattering fraction produces a rise of PD with energy, see e.g. \cite{Nitindala2025, Tomaru2024}. If the system is symmetric, such scattering would not affect the PA. \citet{Nitindala2025} reported PD values similar to those observed from \mbox{Cyg X-2} at inclinations below 70\degr\ for relatively narrow equatorial winds, if the original source of the scattered photons is somewhat beamed along the disk, as is expected for the SL emission. A partly ionized wind illuminated by a central source has been suggested to explain a red-skewed Fe line observed from \mbox{Cyg X-2} \citep{Shaposhnikov2009}. Observations of the source with high-resolution spectroscopic instruments, such as the X-Ray Imaging and Spectroscopy Mission (XRISM), could provide more insight to the existence and properties of a wind in \mbox{Cyg X-2}.

The results from this new observation of \mbox{Cyg X-2} highlight that X-ray polarization provides a powerful tool to understand the nature and geometry of the accretion flow in WMNS binaries. The Comptonized emission from typical BL/SL geometries even when combined with standard reflection models is not predicted to produce the high polarization polarization detected in the HB and the increasing trend of the PD with energy. Therefore, new models are required. Future X-ray spectropolarimetric observations of this source and other similar WMNB binaries, along with improved spectropolarimetric modeling \citep{Nitindala2025,Podgorny2025} are crucial to fully understand the accretion geometry of these systems and the X-ray emission processes.

\begin{acknowledgments}
AG, SB, GM, and FU acknowledge financial support by the Italian Space Agency (Agenzia Spaziale Italiana, ASI) through the contract ASI-INAF-2022-19-HH.0. FC acknowledges financial support by the Istituto Nazionale di Astrofisica (INAF) grant 1.05.23.05.06: ``Spin and Geometry in accreting X-ray binaries: The first multi frequency spectro-polarimetric campaign''. SF and AT acknowledge financial support by the INAF grant 1.05.24.02.04: ``A multi frequency spectro-polarimetric campaign to explore spin and geometry in Low Mass X-ray Binaries''. SF, GM and PS have also been supported by the project PRIN 2022 - 2022LWPEXW - ``An X-ray view of compact objects in polarized light'', CUP C53D23001180006. M.N. is a Fonds de Recherche du Québec – Nature et Technologies (FRQNT) postdoctoral fellow. AS acknowledges the support of the Jenny and Antti Wihuri Foundation (grant no. 00240331).
% Please do not edit the standard acknowledgements for IXPE and NuSTAR
This work reports observations obtained with the Imaging X-ray Polarimetry Explorer (IXPE), a joint US (NASA) and Italian (ASI) mission, led by Marshall Space Flight Center (MSFC). The research uses data products provided by the IXPE Science Operations Center (MSFC), using algorithms developed by the IXPE Collaboration (MSFC, Istituto Nazionale di Astrofisica - INAF, Istituto Nazionale di Fisica Nucleare - INFN, ASI Space Science Data Center - SSDC), and distributed by the High-Energy Astrophysics Science Archive Research Center (HEASARC). This research has made use of data from the NuSTAR mission, a project led by the California Institute of Technology, managed by the Jet Propulsion Laboratory, and funded by NASA. Data analysis was performed using the NuSTAR Data Analysis Software (NuSTARDAS), jointly developed by the ASI Science Data Center (SSDC, Italy) and the California Institute of Technology (USA).
\end{acknowledgments}

\bibliography{nslmxb}{}
\bibliographystyle{aasjournalv7}

\end{document}